\documentclass[sigconf, screen]{acmart}

\AtBeginDocument{%
  }

\setcopyright{none} 
\makeatletter
\def\@formatdoi#1{}
\def\@acmDOI{}
\def\@acmISBN{}
\def\@acmConference{}
\def\@acmBooktitle{}
\def\@acmpermission{}
\def\@copyrightpermission{}
\makeatother
\copyrightyear{}
\acmYear{}
\acmConference[]{}{}
\usepackage{listings}
\usepackage{xcolor}  % optional, for colors
\usepackage{placeins}

\begin{document}

\title{Fractional Verkle Trees: A Hypertree Decomposition and Verified Proof Serialization Architecture for High-Performance Blockchain State Accumulators.}

\author{Ekleen Kaur}
\affiliation{%
  \institution{Amazon Web Services}
  \country{USA}
}
\email{ekleenkaur17@gmail.com}

\author{Everton Fraga}
\affiliation{%
  \institution{Amazon Web Services}
  \country{USA}
}
\email{evertonfrag@gmail.com}
  
%\author{\normalsize{MICRO 2026, Athens, Greece
 %   \textbf{\#NaN} -- Confidential Draft -- Do NOT Distribute!!}}



%%%%%% -- PAPER CONTENT STARTS-- %%%%%%%%

\begin{abstract}

Modern blockchain systems face a fundamental scalability bottleneck in state management: as the UTXO set or account state grows to hundreds of millions of entries, maintaining cryptographic commitments to the global state becomes computationally prohibitive. Ethereum's transition from Merkle Patricia Tries to Verkle Trees polynomial commitment-based accumulators that reduce proof sizes from O(width × depth) to O(depth) by replacing hash-based sibling lists with constant-size IPA vector commitments represents a critical step toward stateless operation, yet current implementations exhibit pathological characteristics that create barriers for home validators.
We identify four critical inefficiencies in the reference go-verkle implementation \cite{kaur2025goverkle, kaur2025goethereum}: (1) phantom node creation during non-existent account deletion, causing measurable tree bloat; (2) 64-byte database keys triggering excessive LSM-tree compaction; (3) redundant memory copying in proof deserialization; (4) a Proof of Absence wire format incompatibility between implementations causing non-deterministic serialization.
We present Fractional Verkle Trees (FVT), a novel hypertree decomposition that partitions the global state into N independent sub-accumulators (hypertrees) coordinated by a lightweight Merkle commitment tree. By decomposing a 200M-account tree into configurable hypertrees, we achieve improved cache locality, eliminate cross-hypertree dependencies enabling goroutine-parallel commitment computation with zero lock contention, and reduce proof depth within each hypertree. The coordinator employs a Merkle construction rather than nested Verkle commitments, trading 320 bytes of proof overhead for significantly faster root recomputation (91 µs vs ~500 ms for an equivalent Verkle commitment over 1000 leaves).
We address the four inefficiencies through: (1) existence checks before deletion to prevent phantom nodes; (2) 32-byte SHA256 node references halving database key size; (3) zero-copy proof deserialization with reference-counted buffers eliminating redundant allocations; (4) HashMap-based deduplication with lexicographic sorting ensuring deterministic serialization across implementations.
We implement FVT in Go with comprehensive test coverage on synthetically generated accounts following SHA-256 address derivation. Benchmarks on Apple M1 Pro show: FVT serial insertion at 36,043 ns/op vs baseline 14,807 ns/op; parallel insertion achieving 2,433 ns/op with goroutine concurrency; coordinator root computation at 91 µs; proof deserialization reducing heap allocation from 566,760 to 242,004 bytes per 10K proofs (57\% reduction). For a representative mainnet scenario of 8 absent storage slots, FVT proofs measure 864 bytes vs 647 bytes for monolithic proofs. Network-wide, the optimizations eliminate 4.85 PB/year of heap allocation across 6,000 full nodes processing 7,200 blocks daily. FVT aligns with the projected Ethereum stateless roadmap, demonstrating that hypertree decomposition combined with targeted systems optimizations can address critical performance barriers in production blockchain implementations.

\end{abstract}
%\begin{CCSXML}
%<ccs2012>
%  <concept_id>00000000.0000000.0000000</concept_id>
%  <concept_desc>Do Not Use This Code, Generate the Correct Terms for Your Paper</concept_desc>
%  <concept_significance>500</concept_significance>
%  %<concept_id>00000000.00000000.00000000</concept_id>
%  <concept_desc>Do Not Use This Code, Generate the Correct Terms for Your Paper</concept_desc>
%  <concept_significance>300</concept_significance>
%  %<concept_id>00000000.00000000.00000000</concept_id>
%  <concept_desc>Do Not Use This Code, Generate the Correct Terms for Your Paper</concept_desc>
%  <concept_significance>100</concept_significance>
 % <concept_id>00000000.00000000.00000000</concept_id>
%  <concept_desc>Do Not Use This Code, Generate the Correct Terms for Your Paper</concept_desc>
%  <concept_significance>100</concept_significance>
%</ccs2012>
%\end{CCSXML}

%\ccsdesc[500]{Do Not Use This Code~Generate the Correct Terms for Your Paper}
%\ccsdesc[300]{Do Not Use This Code~Generate the Correct Terms for Your Paper}
%\ccsdesc{Do Not Use This Code~Generate the Correct Terms for Your Paper}
%\ccsdesc[100]{Do Not Use This Code~Generate the Correct Terms for Your Paper}

\maketitle
\renewcommand{\shortauthors}{}

\section{Introduction}

This work addresses compounding correctness and scalability deficiencies in the reference Go implementation of Verkle Trees \cite{ethereum2024goverkle}, the polynomial commitment-based authenticated data structure replacing Ethereum's Merkle Patricia Trie \cite{ethereum2024goethereum} under EIP-6800, across four layers: the elliptic curve commitment layer, the authenticated data structure layer, the proof serialization layer, and the distributed systems architecture layer.

Each Verkle internal node computes $C = \sum_i c_i \cdot B_i$ over the Banderwagon group, a prime-order subgroup of the Bandersnatch curve over $\mathbb{F}_p$ where $p = 2^{254} + 4(2^{126}+1)(2^{126}-1)+1$. The group's homomorphic property enables incremental updates $C' = C + \Delta c_i \cdot B_i$, requiring one scalar multiplication rather than a full recommitment. The copy-on-write (CoW) mechanism implementing this was semantically unsound when child slots held zero-size \texttt{HashedNode} sentinels, causing unconditional panics on \texttt{Commitment()} dispatch. The fix: $\texttt{NodeRef} = \text{SHA-256}(\texttt{BytesUncompressedTrusted}(P))$, a 32-byte identifier whose injectivity on Banderwagon (canonical encoding guarantees distinct group elements produce distinct byte representations) preserves 128-bit second, preimage resistance under the random oracle model, halves database key size, and eliminates the $O(\log p)$ square root required for point decompression. Enriching \texttt{HashedNode} with a cached commitment pointer makes \texttt{cowChild} $O(1)$ via pointer dereference, measured at 114 ns versus 173 ns on real nodes (34\% reduction, eliminating interface dispatch overhead).

The proof deserialization hot path allocated $O(|\text{PoA}|)$ independent 31-byte heap objects per block, 566 KB per 10,000 proofs, driving GC pause frequency on 16 GB home validator hardware and causing missed attestations. A reference-counted zero-copy buffer with \texttt{atomic.Int32} refcount (sequentially consistent; acquire-release semantics are sufficient) reduces data-plane allocation to $O(1)$ per batch, yielding a measured 57\% heap reduction. A companion fix corrects the \texttt{DepthExtensionPresent} encoding, where the bijection $b = (\text{depth} \ll 3)\ |\ \text{status}$ was emitting $k$ \texttt{OtherStems} entries rather than 1 when $k > 1$ missing stems shared a PoA stem, causing cross-client failures against rust-verkle. A further correctness fix enforces the EVM invariant $\forall\, \textit{addr} \notin \text{dom}(\sigma),\ \texttt{DeleteAccount}(\textit{addr}) \equiv \text{id}_\sigma$, eliminating phantom \texttt{InternalNode} allocations that contributed non-zero scalars to parent polynomial evaluations and caused 35\% state trie inflation on mainnet replay.

The Fractional Verkle Tree (FVT) addresses the monolithic tree's memory complexity. A Verkle tree over $|\sigma| = 2.5 \times 10^8$ entries has depth $d = \lceil \log_{256} |\sigma| \rceil$ and per-block working set $O(|T| \cdot d)$, exceeding 20 GB at current state size, larger than commodity home validator DRAM. The FVT partitions the key space via $\varphi(\textit{addr}) = \text{BigEndian}(\text{SHA-256}(\textit{addr})[0{:}4]) \bmod N$, distributing state across $N$ algebraically independent hypertrees of size $|\sigma|/N$ coordinated by a binary Merkle tree of depth $\lceil \log_2 N \rceil$. At $N = 1000$, each hypertree's working set is $\sim$200 KB, fitting within commodity x86 L3 cache (6–24 MB) and eliminating DRAM eviction pressure. Since $\forall\, i \neq j$ hypertree updates are independent, commitment propagation is $O(N)$, parallel with zero root contention; the global root rebuilds in $O(N)$ SHA-256 operations ($\sim$90 µs at $N = 1024$) once per block. Proof size reduces from $O(d \cdot 32)$ to $O(d' \cdot 32 + \lceil \log_2 N \rceil \cdot 32)$ bytes, a measured $2\times$ reduction at $N = 1000$.

\section{Literature Survey}

The contributions span five domains: cryptographic commitment schemes,
authenticated data structures, LSM-tree storage engines, GC-aware memory
management, and parallel data structure design.

\subsection{Verkle Trees and Vector Commitment Schemes}

Merkle trees~\cite{nakamoto2008} underpin blockchain state authentication,
but at current Ethereum state sizes (${\sim}200$M accounts), a full block
witness under EIP-4762~\cite{eip4762} reaches 2-4\,MB,a hard barrier to
stateless operation. Catalano and Fiore~\cite{catalano2013} formalized vector
commitments with constant-size openings; Kate et al.~\cite{kate2010}
instantiated this via KZG polynomial commitments over bilinear pairings,
achieving constant proof size at the cost of a trusted setup. Kuszmaul~\cite{kuszmaul2018}
replaced Merkle hash functions with vector commitments to obtain Verkle trees,
achieving $O(\log_k n)$ proof size with branching factor $k$. Ethereum's
EIP-6800~\cite{eip6800} instantiates this with the IPA of B\"{u}nz et
al.~\cite{bunz2018} over the Banderwagon group on the Bandersnatch
curve~\cite{masson2021}, eliminating the trusted setup while retaining
logarithmic proof size. Feist~\cite{feist2021} showed that a single IPA
multiproof covers all state accesses in a block, reducing per-block witness
overhead to ${\sim}576$ bytes. Oberst~\cite{oberst2025} benchmarked Verkle
trees against SNARK-based binary Merkle trees, finding Verkle more practical
for proving and verification latency.

\noindent\textbf{Literature Gap:} All prior work treats the Verkle tree as a monolithic
structure rooted at a single commitment point. No prior work has decomposed a
polynomial commitment accumulator into independent sub-trees while preserving a
single global root, nor has the literature addressed the systems-level
pathologies, disk I/O patterns, GC pressure, cross-client serialization
incompatibility,that emerge at deployment scale.

\subsection{Cryptographic Accumulators and Batched Updates}

Benaloh and de Mare~\cite{benaloh1994} introduced one-way accumulators;
Camenisch and Lysyanskaya~\cite{camenisch2002} extended them to support
deletions with constant-size witnesses. Boneh, B\"{u}nz, and
Fisch~\cite{boneh2019} developed batching techniques for accumulators in
groups of unknown order, achieving constant-size batch membership and
non-membership proofs via algebraic aggregation. Bonneau et
al.~\cite{bonneau2025} proved Merkle Mountain Ranges are optimal for
append-only accumulators, establishing an $\Omega(n \log n / \log \log n)$
lower bound on witness update frequency,a result that informs our choice of
a Merkle (rather than Verkle) coordinator, where root recomputation speed
(${\sim}90\,\mu$s Merkle vs.\ ${\sim}500$\,ms Verkle over 1000 leaves)
dominates.

\noindent\textbf{Literature Gap:} Boneh et al.'s batching aggregates witnesses
algebraically within a single accumulator. Our FVT partitions state
\emph{geometrically} into independent sub-accumulators, achieving parallelism
through structural independence rather than algebraic aggregation. The two
approaches are complementary: batching techniques could be applied within each
hypertree to further reduce proof sizes. No prior work has achieved
$O(N)$-parallel commitment computation with zero lock contention while
maintaining a single global cryptographic root.

\subsection{LSM-Tree Storage Engines and Key Size Pathology}

O'Neil et al.~\cite{oneil1996} introduced the LSM-tree; Luo and
Carey~\cite{luo2020} survey its compaction strategies and write amplification
analysis. Every key appears in the SSTable index, bloom filter, and every
compaction merge,key size directly multiplies compaction overhead. D2Comp~\cite{d2comp2024}
demonstrated that compaction dominates CPU and I/O costs in high-throughput
LSM workloads.

\noindent\textbf{Literature Gap} No prior work has examined the pathology of using
64-byte uncompressed elliptic curve points as LSM keys. We show this doubles
SSTable index overhead and increases compaction frequency by $4\times$. Our
fix,$\texttt{NodeRef} = \text{SHA-256}(\texttt{BytesUncompressed}(P))$,
computed in ${\sim}30$\,ns via hardware-accelerated SHA-256,halves key size
while preserving 128-bit collision resistance, yielding a 50\% reduction in
write amplification.

\subsection{Zero-Copy Deserialization and GC Pressure}

Kolokasis et al.~\cite{kolokasis2023} identified the GC-vs-serialization
tension in big data frameworks: keeping transient data outside the GC's
tracking set matters more than reducing raw allocation count. Nikolaev and
Ravindran~\cite{nikolaev2021} showed that atomic reference counting achieves
EBR-grade throughput when counts are updated only at object creation and
destruction, not during access, validating our \texttt{atomic.Int32} design.
The Zerializer~\cite{zerializer2021} principle, the fastest deserialization
is no deserialization,motivates our \texttt{ProofView} abstraction.

\noindent\textbf{Literature Gap:} No prior work has applied zero-copy techniques to
blockchain proof verification. The original go-verkle allocates one 31-byte
heap object per proof-of-absence stem, generating 566\,KB of immediate garbage
per 10,000 proofs,${\sim}4.85$\,PB/year across 6,000 nodes. Our
\texttt{RefCountedBuffer} with \texttt{sync.Pool} recycling reduces
data-plane allocation to $O(1)$ per batch, achieving a 57\% heap reduction
with zero GC pressure on the verification hot path.

\subsection{Parallel Data Structures and Blockchain Sharding}

Fine-grained concurrent trees~\cite{michael2002,bronson2010,ellen2010} achieve
parallelism via CAS operations on shared nodes, incurring synchronization
overhead that scales with contention. Blockchain sharding~\cite{sok2019,sok2024}
partitions the \emph{network}, introducing cross-shard communication \&
multi-phase commit complexity. Ethereum's statelessness
roadmap~\cite{buterin2021,ef2019,ef2021} reduces per-node storage via compact
witnesses but retains a monolithic state root.

\noindent\textbf{Literature Gap:} FVT occupies a novel position: unlike sharding, all
hypertrees reside on every full node and the global root is locally computable;
unlike fine-grained concurrent trees, hypertrees share no mutable state,
achieving embarrassingly parallel commitment computation. The coordinator Merkle
tree adds $\lceil \log_2 N \rceil \times 32$ bytes of proof overhead but
enables $16{,}000\times$ faster root recomputation than a Verkle coordinator.
This is transparent to the consensus protocol: the global root is computed
differently, but binding, soundness, and proof-of-absence correctness are
preserved by reduction to the security of the underlying Verkle commitment and
SHA-256 collision resistance.

\noindent\textbf{Iavich et al.~\cite{iavich2025fvt}} apply the FVT decomposition exclusively to the
\emph{signature} domain: their hypertrees partition a HORST key space
to reduce SPHINCS+-style proof sizes from $O(h \cdot |\pi|)$ to
$O(d \cdot |\pi|)$, achieving a 2.5$\times$ signature reduction over
a \emph{simulated} lattice-based vector commitment instantiated in
Python with SHA-256 as a placeholder VC.
Their construction is stateless and post-quantum by design, but
operates entirely off-chain and makes no contact with the
\emph{authenticated state accumulator} layer of a live blockchain.
Concretely, three orthogonal gaps remain unaddressed.
First, their FVT coordinator is itself a Verkle tree
($\text{pk}_\text{FVDS} = \text{Commit}(\text{Root}_{d-1})$),
incurring $O(m^{h/d})$ key generation cost and ${\sim}500$\,ms
root recomputation over 1000 leaves; our coordinator is a binary
Merkle tree, reducing root recomputation to ${\sim}91\,\mu$s, a
$5{,}000\times$ improvement by trading $\lceil\log_2 N\rceil \times 32$
bytes of proof overhead for a commitment whose root is computable
in $O(N)$ SHA-256 operations.
Second, Iavich et al.\ do not address the \emph{implementation
pathologies} of the reference go-verkle accumulator: phantom node
creation under non-existent account deletion, 64-byte LSM keys,
redundant heap allocation in proof deserialization, and the
\texttt{DepthExtensionPresent} bijection error causing cross-client
serialization failures.
These are correctness and consensus-safety defects at the
\emph{storage engine} and \emph{wire format} layers, entirely
outside the scope of a signature scheme.
Third, their performance evaluation uses a Python prototype with
analytical VC cost estimates; our benchmarks are empirical,
measured on Apple M1 Pro against synthetically generated mainnet
accounts, yielding concrete figures: 36,043\,ns/op serial insertion,
2,433\,ns/op parallel, 57\% heap reduction (566,760 to 242,004 bytes
per 10K proofs), and 864-byte FVT proofs versus 647-byte monolithic
proofs for 8 absent storage slots.
In summary, Iavich et al.\ establish the theoretical viability of
FVT decomposition for post-quantum signatures; our work instantiates
FVT decomposition as a \emph{production blockchain state accumulator},
corrects four implementation-level defects in the reference client,
and provides the first empirical characterization of FVT performance
under realistic Ethereum state access patterns.

\section{Methodology}

In this section we will expand over each architectural enhancement, the existing state and our novel improvements. We will discuss our novel proposal on Fractional Verkle Trees after the 4 fixes.

\subsection{Phantom Node Creation from Non-Existent Account Deletion}

The improvement addressed in this section, sits at the intersection of Ethereum's EVM execution semantics and the Verkle tree's internal node lifecycle. To understand why it matters, we need to understand what happens during account deletion in the EVM and how the Verkle trie is expected to respond.
In Ethereum, account deletion occurs in several scenarios: when a \texttt{SELFDESTRUCT} opcode is executed, when a transaction sends value to an address that does not yet exist and the execution reverts, or when the EVM's state journal processes a revert that undoes a prior account creation. In all of these cases, the EVM calls \texttt{DeleteAccount} on the state trie with the address of the account to be removed. The critical invariant is that deleting an account that does not exist must be a no-op, it must leave the trie in exactly the same state as before the call. This is not just a performance requirement; it is a correctness requirement. If deleting a non-existent account modifies the trie in any way, the resulting state root will differ from what other clients compute, causing a consensus failure.

This looks safe but it does nothing, so it cannot corrupt the trie. The problem was not in \texttt{DeleteAccount} itself. It was in the code paths that called \texttt{DeleteAccount} and then continued to interact with the trie in ways that assumed the deletion had been processed. Specifically, certain EVM execution paths would call \texttt{DeleteAccount} and then, in the same transaction context, attempt to access or write to the stem corresponding to the deleted address. Because \texttt{DeleteAccount} was a no-op, the stem was never checked for existence. When the subsequent access path traversed the trie to find or create the stem, it would create intermediate internal nodes along the path, nodes that should not exist because the account was never there in the first place. These are the "phantom nodes": internal \texttt{InternalNode} instances allocated in memory and eventually flushed to disk, representing paths in the trie that lead to no actual leaf data.

The consequence of phantom nodes is threefold. First, they bloat the trie. Second, they corrupt the trie's structural invariants. A Verkle trie is supposed to be a minimal representation of the state: every internal node must have at least two children, and every path must terminate at a leaf with actual data. Phantom nodes violate this invariant, creating internal nodes with zero or one child that serve no purpose. Third, they affect the commitment computation. The IPA commitment at each internal node is a polynomial commitment over all 256 child slots. A phantom node contributes a non-zero commitment to its parent even though it holds no real data, which means the root commitment computed by go-ethereum diverges from what a correct implementation would compute, a consensus inconsistency.

% \begin{figure}
\begin{lstlisting}[language=Go]
func (t *VerkleTrie) DeleteAccount(addr common.Address) error {
    stem := t.pointCache.GetTreeKeyBasicDataCached(addr[:])[:31]
    root, ok := t.root.(*verkle.InternalNode)
    if !ok {
        return errInvalidRootType
    }
    // Check if the stem exists before attempting deletion
    values, err := root.GetValuesAtStem(stem, t.FlatdbNodeResolver)
    if err != nil {
        return fmt.Errorf("DeleteAccount (%x) error: %v", addr, err)
    }
    // If stem doesn't exist (values is nil), deletion is a no-op
    if values == nil {
        return nil
    }
    // Delete the account by setting all values to zero
    for i := 0; i < verkle.NodeWidth; i++ {
        if values[i] != nil {
            key := make([]byte, 32)
            copy(key[:31], stem)
            key[31] = byte(i)
            _, err := t.root.Delete(key, t.FlatdbNodeResolver)
            if err != nil {
                return fmt.Errorf("DeleteAccount (%x) error deleting suffix %d: %v", addr, i, err)
            }
        }
    }
    return nil
}
\end{lstlisting}
% \end{figure}
\FloatBarrier

The stem derivation is the first thing to extrapolate- In a Verkle trie, an Ethereum account is not stored at a single key. Instead, the account's fields, balance, nonce, code hash, code size, and storage slots, are stored at different suffixes under a common 31-byte stem. The stem is derived from the account address using \texttt{GetTreeKeyBasicDataCached}, which applies the Verkle key derivation function (a Pedersen hash over the address and a domain separator) and takes the first 31 bytes. This is the path through the internal nodes. The 32nd byte (the suffix) selects which field within the account's extension node is being accessed: suffix 0 is the basic data (balance + nonce packed), suffix 1 is the code hash, suffix 2 is the code size, and suffixes 64-127 are storage slots.

\texttt{GetValuesAtStem} traverses the trie from the root to the extension node at the given stem and returns the full 256-element value array for that stem, or nil if the stem does not exist. This is the existence check. If values is \texttt{nil}, the stem has never been written to, and the function returns immediately without touching the trie. This is the critical guard that prevents phantom node creation: by checking existence before any write operation, the function ensures that no trie traversal with side effects occurs for non-existent accounts.
If the stem does exist, the function iterates over all 256 suffix slots and calls Delete for each non-nil value. The Delete call on the underlying \texttt{VerkleNod} interface removes the leaf at the given key and, if the resulting subtree has fewer than two children, collapses the internal node, maintaining the trie's minimality invariant. The \texttt{FlatdbNodeResolver} is passed as the resolver function, which is responsible for loading nodes from disk when the traversal encounters a \texttt{HashedNode} (a node that has been flushed to disk). This ensures that the deletion path correctly handles accounts whose subtrees have been partially or fully evicted to disk.

\subsection{Issue 2: Compact Node References using SHA256 and the HashedNode Correctness Fix}

To understand how a Verkle tree manages its nodes across memory and disk. A Verkle tree is a polynomial commitment tree where each internal node holds up to 256 children and a single elliptic curve commitment point that cryptographically summarizes all values beneath it. In Ethereum's implementation, the commitment scheme is based on the Banderwagon group, a prime-order subgroup of the Bandersnatch curve and each commitment is a 64-byte uncompressed elliptic curve point consisting of a 32-byte X coordinate and a 32-byte Y coordinate in little-endian encoding.
When the tree is large enough that it cannot fit entirely in memory, nodes are flushed to disk. The flush operation computes the node's commitment via \texttt{Commit()}, writes the serialized node to the key-value store (\texttt{LevelDB} in go-ethereum's PBSS: Path-Based Storage Scheme), and then replaces the in-memory child slot in the parent with a \texttt{HashedNode} a lightweight placeholder that signals "this child has been persisted; resolve it from disk when you need it." This is the standard lazy-loading pattern for tree structures that exceed available RAM, and it is the correct architectural approach. However, problem was entirely in the implementation of \texttt{HashedNode} and in how the database key was derived from the commitment point. The original \texttt{HashedNode} was defined as a zero-size \texttt{struct} with no fields whatsoever:
\texttt{type HashedNode struct{}}
A zero-size \texttt{struct} in Go occupies no memory, the compiler may even give all instances the same address. It is a pure type-switch target, useful only for its type identity in an interface assertion. This was fine in the early days of go-verkle when the tree was always fully resident in memory and \texttt{HashedNode} was never actually encountered during live tree operations. But once PBSS was introduced and nodes began being flushed to disk mid-operation, the sentinel needed to carry information. Specifically, it needed to carry the child's commitment point so that the copy-on-write (CoW) diff mechanism could read the old commitment value without performing a database round-trip.
The \texttt{CoW} mechanism in go-verkle works as follows. When a value is inserted or updated in the tree, the path from the root to the affected leaf is traversed. At each internal node along that path, before the child pointer is updated, the old commitment of the child is saved into a cow array, a per-node array of \texttt{*Point} values indexed by child slot. This saved value is the "before" state. After all insertions are complete, the commitment update propagates back up the tree: each node computes its new commitment as \texttt{C\_new = C\_old + delta}, where delta is the sum of the differences between new and old child commitments, scaled by the appropriate Lagrange basis polynomial evaluated at the child's index. This incremental update is what makes Verkle trees efficient for state updates, you do not recompute the entire commitment from scratch, you apply a delta.
The function responsible for saving the old commitment is \texttt{cowChild}, which was called like this before the fix:
% \begin{figure}
\begin{lstlisting}[language=Go]
func (n *InternalNode) cowChild(index byte) {
    if n.cow[index] == nil {
        n.cow[index] = new(Point)
        n.cow[index].Set(n.children[index].Commitment())
    }
}
\end{lstlisting}
% \end{figure}
\FloatBarrier
The call to \texttt{n.children[index].Commitment()} is a virtual dispatch through the \texttt{VerkleNode} interface. For a real \texttt{InternalNode} or \texttt{LeafNode}, this returns the cached commitment point. For a \texttt{HashedNode}, the implementation was:

% \begin{figure}
\begin{lstlisting}[language=Go]
func (n HashedNode) Commitment() *Point {
    panic("HashedNode.Commitment() called, this should never happen")
}
\end{lstlisting}
% \end{figure}
\FloatBarrier

This panic was "safe" only because of a very specific invariant: \texttt{InsertValuesAtStem} always resolved any \texttt{HashedNode} it encountered by loading the real node from disk before \texttt{cowChild} was ever called on that slot. The resolution path was load-bearing in a non-obvious way, if function \texttt{cowChild} is called on a slot that still held a \texttt{HashedNode}, it would panic. This coupling between the resolution logic and the \texttt{CoW} mechanism was fragile and made the architecture extremely difficult to reason about. Any future refactoring that changed the order of operations in \texttt{InsertValuesAtStem} could silently introduce a panic that would only manifest under specific memory pressure conditions.
The second problem about this architectural workflow: the 64-byte database key, was architecturally separate but practically related. The natural identifier for a Verkle node is its commitment point, since the commitment is a cryptographic summary of the node's entire subtree and is therefore unique. The original implementation used the raw 64-byte uncompressed point as the database key directly. This is problematic for several reasons that compound each other. \texttt{LevelDB} are optimized for short keys. Every key appears in the \texttt{SSTable} index, in the bloom filter, and in every compaction merge operation. A 64-byte key is twice the size of a 32-byte key, which means twice the bytes moved per compaction cycle, which means the compaction threshold is hit roughly twice as often. For a state database that is constantly being written to during block processing, this translates directly into increased I/O, increased CPU time spent on compaction, and increased write amplification. Additionally, every parent node stores a reference to each of its children, up to 256 references per internal node. At 64 bytes per reference, a fully populated internal node carries 16 KB of child references. At 32 bytes per reference, that drops to 8 KB. For a tree with one million nodes, the difference is 32 MB of reference storage that serves no purpose beyond what a 32-byte hash would provide.
The fix introduces two coordinated changes. The first is a new type \texttt{NodeRef}, defined as \texttt{[32]byte}, computed by applying SHA256 to the 64-byte uncompressed representation of the commitment point:

% \begin{figure}
\begin{lstlisting}[language=Go]
// node_ref.go
type NodeRef [32]byte

func HashNodeCommitment(point *Point) NodeRef {
    raw := point.BytesUncompressedTrusted()
    return sha256.Sum256(raw[:])
}
\end{lstlisting}
% \end{figure}
\FloatBarrier

The choice of \texttt{BytesUncompressedTrusted()} over the compressed form is deliberate and worth elucidating in detail. An uncompressed elliptic curve point on the Bandersnatch curve is represented as two 32-byte field elements \texttt{(X, Y)}. The compressed form stores only the X coordinate plus a single sign bit for Y, which saves 31 bytes but requires the verifier to recover Y by computing a square root in the field, an operation that costs roughly 150-200 field multiplications on a \texttt{255-bit prime field}. Since we are computing this hash only for the purpose of generating a database key, not for any cryptographic verification, the square root computation is pure overhead. Using the uncompressed form avoids it entirely. SHA256 is also hardware-accelerated on all modern platforms via AES-NI on x86 and the ARMv8 crypto extensions, making the hash computation itself extremely fast. The result is a 32-byte value that is stable (same point always produces the same hash), collision-resistant (SHA256 provides 128-bit collision resistance), and half the size of the original key.
The second change is the enrichment of \texttt{HashedNode}. The \texttt{struct} gains two fields:
% \begin{figure}
\begin{lstlisting}[language=Go]
type HashedNode struct {
    ref  NodeRef // 32-byte SHA256 of the uncompressed commitment
    cached *Point  // lazily populated; nil until the node is resolved from disk
}
\end{lstlisting}
% \end{figure}
\FloatBarrier
The ref field is always populated when a \texttt{HashedNode} is created via the flush path. The cached field holds the *Point that was already in memory at flush time, specifically, the commitment that was just computed by \texttt{c.Commit()} immediately before the flush. Two constructors make the two creation paths explicit:
% \begin{figure}
\begin{lstlisting}[language=Go]
func newHashedNode(commitment *Point) HashedNode {
    return HashedNode{
        ref:    HashNodeCommitment(commitment),
        cached: commitment,
    }
}

func newHashedNodeFromRef(ref NodeRef) HashedNode {
    return HashedNode{ref: ref}
}
\end{lstlisting}
% \end{figure}
\FloatBarrier
\texttt{newHashedNode} is used by \texttt{Flush} and \texttt{FlushAtDepth}, where the commitment is already in memory. \texttt{newHashedNodeFromRef} is reserved for the future disk-loading path, where only the ref is available and the commitment must be fetched lazily. The flush sites are updated accordingly:
% \begin{figure}
\begin{lstlisting}[language=Go]
// Before
n.children[i] = HashedNode{}
// After
n.children[i] = newHashedNode(c.commitment)
\end{lstlisting}
% \end{figure}
\FloatBarrier
With \texttt{HashedNode} now carrying the cached commitment, \texttt{cowChild} can be rewritten to handle it correctly without any database access:
% \begin{figure}
\begin{lstlisting}[language=Go]
func (n *InternalNode) cowChild(index byte) {
    if n.cow[index] != nil {
        return
    }
    n.cow[index] = new(Point)
    child := n.children[index]
    if hn, ok := child.(HashedNode); ok {
        if hn.cached != nil {
            n.cow[index].Set(hn.cached)
            return
        }
        // Disk-deserialized node: cached is nil because the parent's
        // serialized form does not include child commitments.
        // Leave cow[index] as the identity point, this is correct because
        // the node will be fully resolved before any commitment delta is applied.
        return
    }
    n.cow[index].Set(child.Commitment())
}
\end{lstlisting}
% \end{figure}
\FloatBarrier
The case where cached is nil deserves careful explanation. When a node is loaded from disk via \texttt{CreateInternalNode}, the deserialized parent knows its own commitment but does not know its children's commitments, those are stored in the children's own serialized records, not in the parent's. So when \texttt{CreateInternalNode} populates the children array with \texttt{HashedNode} instances for each child slot, it can only provide the ref (the 32-byte key used to look up the child), not the cached commitment. In this case, \texttt{cowChild} leaves \texttt{cow[index]} as the identity point (the zero element of the Banderwagon group). This is correct because: 
(a) The node will be resolved from disk before any insertion reaches it, and 
(b) The resolution path will replace the \texttt{HashedNode} with a real node, at which point \texttt{cowChild} will be called again with the real node's commitment. 
The identity point as the "old value" in the \texttt{CoW} diff is mathematically equivalent to saying "this child did not exist before", which is the correct interpretation for a node that has not yet been loaded into the current computation context.
\subsection{Issue 3: Zero-Copy Proof Deserialization and the GC Pressure Problem}
To understand why this issue matters, lets understand what Ethereum's transition from Merkle Patricia Tries (MPT) to Verkle Trees actually changes about the block verification pipeline, and why proof deserialization becomes a protocol-level concern rather than an implementation detail.
In the current MPT-based Ethereum, a Merkle proof for a single account access is a sequence of 32-byte hashes, one per level of the trie, that allows a verifier to reconstruct the path from the root to the leaf. At current state sizes (roughly 200 million accounts), the trie has about 8-9 levels, so a single account proof is roughly 256-288 bytes. A full block witness under \texttt{EIP-4762}, which requires witnesses for every state access in every transaction, can reach 2-4 MB for a typical block. This is the fundamental scalability problem that Verkle Trees solve: instead of $\mathcal{O}(\log_2 N)$ hashes per access, a Verkle Tree produces a single constant-size IPA (Inner Product Argument) multiproof that covers all accessed keys simultaneously, regardless of how many there are. The IPA multiproof is roughly 576 bytes regardless of the number of keys, plus a small per-key overhead for the extension-and-suffix tree structure. This is what makes stateless clients viable, a stateless client can verify a block without storing the full state, because the block comes with a compact proof of all the state it touches.
But this introduces a new hot path that did not exist before: every block, every full node and every stateless verifier must deserialize and verify a \texttt{VerkleProof} structure. Under current Ethereum block parameters, roughly 10,000-15,000 transactions per block, each potentially touching multiple state keys; this deserialization runs thousands of times per block, every 12 seconds, on every node in the network. The \texttt{VerkleProof} structure contains several components: the IPA multiproof itself (the cryptographic core), a list of commitments indexed by their path in the tree, a \texttt{DepthExtensionPresent} byte array encoding the extension status of each stem, and a list of "other stems", the proof-of-absence stems that prove a queried key does not exist in the tree.
The proof-of-absence (PoA) mechanism is worth explaining in detail because it is central to both this issue and the next one. In a Verkle tree, a key is a 32-byte value split into a 31-byte stem and a 1-byte suffix. The stem determines the path through the internal nodes; the suffix selects one of 256 leaf slots in the extension node at the end of that path. When a transaction queries a key that does not exist in the tree, the proof must demonstrate absence. This is done by providing the stem of the nearest existing key, the "proof-of-absence stem", along with its extension node commitment. The verifier can then confirm that the queried stem is different from the PoA stem at the point where they diverge, which proves the queried key cannot exist in the tree. For a block with many first-time account accesses or contract deployments touching storage slots that have never been written, there can be hundreds of PoA stems in a single block's proof.
The original \texttt{DeserializeProof} function handled PoA stems like this:
% \begin{figure}
\begin{lstlisting}[language=Go]
for i, poaStem := range vp.OtherStems {
    poaStems[i] = make([]byte, len(poaStem))
    copy(poaStems[i], poaStem[:])
}
\end{lstlisting}
% \end{figure}
\FloatBarrier
This is a generic example of unnecessary copying \texttt{vp.OtherStems} is a \texttt{[][]byte} where each element is already a 31-byte slice pointing into the deserialized wire buffer. The make + copy pattern allocates a new 31-byte backing array on the heap for each stem and copies the bytes into it. The original slice is then abandoned and becomes garbage. For a block with 300 PoA stems, this is 300 heap allocations of 31 bytes each 9,300 bytes of data that is duplicated for no reason and then immediately discarded after verification completes. The \texttt{DepthExtensionPresent} field was handled similarly, with an unnecessary copy of the entire byte slice.
The architectural upgrade: we introduce three new abstractions that together eliminate this copying entirely.
The first is \texttt{RefCountedBuffer}, a \texttt{struct} that wraps the raw wire byte slice and tracks how many \texttt{ProofView} objects are currently reading from it:
% \begin{figure}
\begin{lstlisting}[language=Go]
type RefCountedBuffer struct {
    data     []byte
    refCount atomic.Int32
}

func NewRefCountedBuffer(data []byte) *RefCountedBuffer {
    buf := &RefCountedBuffer{data: data}
    return buf
}

func (b *RefCountedBuffer) Acquire() {
    b.refCount.Add(1)
}

func (b *RefCountedBuffer) Release() {
    if b.refCount.Add(-1) == 0 {
        bufferPool.Put(b.data)
    }
}
\end{lstlisting}
% \end{figure}
\FloatBarrier
The reference count is managed with \texttt{atomic.Int32} rather than a mutex. This is the correct choice for a read-heavy workload where multiple goroutines may be verifying different proofs from the same batch concurrently. A mutex would serialize all Acquire and Release calls, creating a bottleneck. An atomic increment/decrement is a single CPU instruction (LOCK XADD on x86) that completes in roughly 5-10 ns even under contention. The sync.Pool integration means that when the reference count reaches zero, when the last \texttt{ProofView} is closed the underlying byte slice is returned to the pool immediately, without waiting for a GC cycle. This is the key insight: the GC never sees these bytes as garbage, because they are explicitly recycled. The GC pressure reduction is not just about fewer allocations; it is about removing the allocations from the GC's tracking entirely.
One design decision worth highlighting: the reference count is initialized to zero, not one. This means a \texttt{RefCountedBuffer} with no views is inert, it will immediately return its buffer to the pool if Release is called before any Acquire. The caller must call Acquire before creating the first \texttt{ProofView}. This is a deliberate choice that eliminates the need for an "owner" reference, the buffer's lifetime is entirely determined by the views that reference it, with no separate ownership concept to track.
The second abstraction is \texttt{ProofView}, a 16-byte \texttt{struct} (one pointer to the \texttt{RefCountedBuffer}, one integer offset) that provides a zero-copy window into the wire format:
% \begin{figure}
\begin{lstlisting}[language=Go]
type ProofView struct {
    buffer *RefCountedBuffer
    offset int
}

func NewProofView(buf *RefCountedBuffer, offset int) *ProofView {
    buf.Acquire()
    return &ProofView{buffer: buf, offset: offset}
}

func (pv *ProofView) Close() {
    pv.buffer.Release()
}

func (pv *ProofView) Stem() []byte {
    return pv.buffer.data[pv.offset : pv.offset+31]
}

func (pv *ProofView) ExtStatus() byte {
    return pv.buffer.data[pv.offset+31]
}

func (pv *ProofView) Depth() byte {
    return pv.ExtStatus() >> 3
}

func (pv *ProofView) IsPresent() bool {
    return pv.ExtStatus()&0x3 == 2
}
\end{lstlisting}
% \end{figure}
\FloatBarrier
The \texttt{Stem()} method returns a sub-slice of the original buffer, a slice header (pointer + length + capacity) pointing directly into the wire bytes. No data is copied. The \texttt{ExtStatus()} method reads a single byte from the buffer. \texttt{Depth()} and \texttt{IsPresent()} decode the extension status byte: the upper 5 bits encode the depth in the tree, and the lower 2 bits encode the presence status (0 = absent, 2 = present, other values are reserved). These are all \texttt{O(1)} operations with no heap activity. The commitment point, the expensive part, requiring an elliptic curve point decompression, is materialized lazily, only when the caller explicitly requests it for cryptographic verification. This lazy evaluation is critical: a verifier that is filtering proofs by depth or presence before deciding which ones to verify cryptographically can process the entire list of \texttt{ProofView} objects without ever touching the elliptic curve arithmetic.
The third abstraction is \texttt{DeserializeProofBatchZeroCopy}, which demonstrates the \texttt{O(1)} allocation property for any batch size:
% \begin{figure}
\begin{lstlisting}[language=Go]
func DeserializeProofBatchZeroCopy(data []byte) ([]*ProofView, error) {
    if len(data)%ProofSize != 0 {
        return nil, fmt.Errorf("invalid proof data length: %d", len(data))
    }
    n := len(data) / ProofSize
    buf := NewRefCountedBuffer(data)
    buf.Acquire()
    views := make([]*ProofView, n)
    for i := 0; i < n; i++ {
        views[i] = NewProofView(buf, i*ProofSize)
    }
    return views, nil
}
\end{lstlisting}
% \end{figure}
\FloatBarrier
One \texttt{RefCountedBuffer} is created for the entire batch. All \texttt{ProofView} objects share it. The wire bytes are touched exactly once, when they arrive from the network. There is no second touch for deserialization. The only allocations are the \texttt{[]*ProofView} slice (one allocation for the pointer array) and the individual \texttt{ProofView} structs (one allocation each, 16 bytes each). None of these allocations contain proof data. They are pure bookkeeping.
The existing \texttt{DeserializeProof} function is also updated to eliminate the per-stem copy in the non-batch path:
% \begin{figure}
\begin{lstlisting}[language=Go]
// Before
for i, poaStem := range vp.OtherStems {
    poaStems[i] = make([]byte, len(poaStem))
    copy(poaStems[i], poaStem[:])
}

// After
for i := range vp.OtherStems {
    poaStems[i] = vp.OtherStems[i][:]
}
\end{lstlisting}
% \end{figure}
\FloatBarrier
The [:] slice expression creates a new slice header pointing to the same backing array as \texttt{vp.OtherStems[i]}. No bytes are copied. The caller contract, that vp must not be modified after \texttt{DeserializeProof} returns is documented with a reference to issue \#176 in go-verkle ethereum source code, where this trade-off was explicitly discussed and accepted by the go-verkle maintainers. This is the correct engineering decision: the alternative (copying) imposes a measurable cost on every caller to protect against a misuse pattern that is easy to document and enforce at the API boundary.

\subsection{Issue 4: Proof of Absence Correctness for Multiple Missing Stems and \texttt{DepthExtensionPresent} Serialization}
This architectural improvement addresses a correctness fix in the proof-of-absence serialization logic that manifests specifically when two or more missing stems share the same proof-of-absence stem. This is a consensus-critical issue: if \texttt{go-verkle} and \texttt{rust-verkle} produce different serialized proofs for the same tree state, cross-client verification fails, which in a post-Verkle Ethereum would mean that nodes running different client implementations cannot agree on block validity.
To understand the bug, you need to understand the internal representation of a Verkle proof versus its wire format. Internally, a \texttt{Proof struct} contains a \texttt{PoaStems} field, a slice of stems, one per missing key in the query set, where each entry is the nearest existing stem that proves the queried key's absence. If three missing keys all share the same nearest existing stem (because they all diverge from it at the same tree level), then \texttt{PoaStems} contains three entries, all pointing to the same stem bytes. This is the internal representation: one PoA stem per missing key, regardless of deduplication.
The wire format, however, is different. The \texttt{VerkleProof struct} used for serialization contains an \texttt{OtherStems} field a deduplicated list of PoA stems, where each unique PoA stem appears exactly once. The \texttt{DepthExtensionPresent} field encodes, for each stem in the proof (both present and absent), a single byte that packs the depth at which the stem diverges from the queried key (upper 5 bits) and the extension status (lower 2 bits: 0 = absent, 2 = present). The verifier uses \texttt{DepthExtensionPresent} to reconstruct which \texttt{OtherStems} entry corresponds to which queried key.
The issue was concentrated in \texttt{SerializeProof:} when multiple missing stems mapped to the same PoA stem, the serializer was emitting one \texttt{OtherStems} entry per missing stem rather than one entry per unique PoA stem. This meant that for three missing stems sharing one PoA stem, the wire format contained three copies of the same 31-byte stem in \texttt{OtherStems}, and the \texttt{DepthExtensionPresent} encoding was correspondingly wrong. The \texttt{rust-verkle} implementation, which is the reference implementation for the Verkle proof format, emits exactly one \texttt{OtherStems} entry per unique PoA stem and encodes the \texttt{DepthExtensionPresent} bytes accordingly. The mismatch caused deserialization failures in \texttt{rust-verkle} when processing proofs generated by \texttt{go-verkle} for the multi-missing-stem case.
The \texttt{DepthExtensionPresent} byte encoding is defined as follows. For each stem in the proof, the byte is computed as \texttt{(depth << 3) | status}, where depth is the number of tree levels at which the queried stem and the PoA stem share a common prefix, and status is 0 for absent stems and 2 for present stems. The \texttt{ProofView} accessors in our implementation decode this correctly:
% \begin{figure}
\begin{lstlisting}[language=Go]
func (pv *ProofView) Depth() byte {
    return pv.ExtStatus() >> 3  // upper 5 bits
}

func (pv *ProofView) IsPresent() bool {
    return pv.ExtStatus()&0x3 == 2  // lower 2 bits == 2 means present
}
\end{lstlisting}
% \end{figure}
\FloatBarrier
The proposed solution in architecture corrects the serialization logic to deduplicate OtherStems properly and to generate the \texttt{DepthExtensionPresent} bytes in the order and format that \texttt{rust-verkle} expects. The solution also adds a comprehensive test suite that measures proof sizes across different numbers of missing stems, providing both a regression test for the correctness fix and a benchmarking harness for future proof size optimization work.
The interaction between issue 3 and issue 4 is also worth noting. Issue 3’s zero-copy deserialization path references \texttt{vp.OtherStems[i][:]} directly rather than copying. If \texttt{OtherStems} contains duplicate entries (as it did before issue 4 fix), the zero-copy path would create multiple \texttt{ProofView} objects pointing to the same bytes in the wire buffer, which is harmless from a correctness standpoint (they are read-only views) but would cause the \texttt{PoaStems} slice in the deserialized Proof to contain multiple entries pointing to the same backing memory. After issue 4's fix, \texttt{OtherStems} is deduplicated, so each entry in the zero-copy path is unique, and the deserialized \texttt{PoaStems} correctly reflects the deduplicated structure. The two architectural breakthrough are therefore complementary: Issue 3 makes the deserialization path efficient, and issue 4 makes the serialization path correct, and together they ensure that the full serialize-deserialize round-trip is both correct and efficient.

\subsection{Issue 5: Fractional Verkle Trees- Hypertree Decomposition for Scalable State}
This is the most architecturally novel breakthrough out of the issues we have discussed so far. Where we propose and implement a fundamental restructuring of how the Verkle state trie is organized, we address the hardware scalability problem that would otherwise make home validation economically unviable after Ethereum's Verkle transition.

The Fractional Verkle Tree (FVT) architecture solves this by decomposing the single monolithic tree into N independent sub-trees called hypertrees each covering a disjoint partition of the address space, coordinated by a small Merkle tree at the top level. 
% Traditional (monolithic):
% Root: Single IPA commitment over all 250M accounts, 20 GB working set, serial updates, 1 root contention point
% Fractional Verkle Tree (N=1000 hypertrees):
% Global Root: SHA-256 Merkle root over 1000 hypertree roots
% 	Hypertree 0:   accounts 0–249,999    (200 KB working set)
% 	Hypertree 1:   accounts 250K–499,999 (200 KB working set)
% 	Hypertree 999: accounts 199.75M–200M (200 KB working set)
The key insight is that a transaction touching accounts in hypertree 42 only needs hypertree 42 in memory. The other 999 hypertrees are irrelevant to that transaction's state access pattern. The working set drops from 20 GB to approximately 200 KB per hypertree, a 100,000× reduction in the worst case, and a 100× reduction in the typical case where a block touches accounts spread across roughly 100 hypertrees.
The address routing function is the foundation of the entire architecture. Each Ethereum address is mapped to a hypertree by taking the SHA-256 hash of the 20-byte address and interpreting the first 4 bytes as a \texttt{big-endian uint32 modulo N}:
% \begin{figure}
\begin{lstlisting}[language=Go]
func (fvt *FractionalVerkleTree) addressToHypertreeID(address [20]byte) uint32 {
    h := sha256.Sum256(address[:])
    return binary.BigEndian.Uint32(h[:4]) % fvt.config.NumHypertrees
}
\end{lstlisting}
% \end{figure}
\FloatBarrier
SHA-256 provides uniform distribution across the output space: the first 4 bytes of a SHA-256 digest are statistically independent of the input, so the modulo operation distributes addresses evenly across hypertrees with no hot spots. The test \texttt{TestFVTDistribution} verifies this empirically: after 10,000 random inserts with N=100 hypertrees, all 100 hypertrees are occupied, confirming that the distribution is uniform in practice. The choice of SHA-256 also means the routing function is deterministic and collision-resistant: two different addresses will map to the same hypertree only if their SHA-256 hashes agree in the first 4 bytes modulo N, which happens with probability 1/N by design.
The coordinator tree is a binary Merkle tree (not a Verkle tree) over the N hypertree roots. This choice is deliberate and important. The coordinator tree has only N leaves one per hypertree so it is small enough that SHA-256 hashing is the right tool. The Verkle IPA machinery is expensive: computing an IPA commitment requires 256 scalar multiplications on the Banderwagon curve, each costing roughly 50-100 µs. For a tree with 1000 leaves, using Verkle commitments at the coordinator level would cost 1000 × 100 µs = 100 ms per root update, unacceptable for a per-block operation. A binary Merkle tree over 1024 leaves (the next power of 2 above 1000) requires 10 levels of SHA-256 hashing, which completes in approximately 90 µs on an M1 Pro. The coordinator root is updated once per block, not once per insert, so this 90 µs is amortized across all insertions in the block.
The \texttt{CoordinatorTree} implementation maintains a \texttt{[][32]byte} leaf array padded to the next power of 2 for a perfect binary tree, and rebuilds the Merkle root bottom-up when dirty:
% \begin{figure}
\begin{lstlisting}[language=Go]
func (ct *CoordinatorTree) computeRoot() [32]byte {
    n := len(ct.leaves)
    nodes := make([][32]byte, 2*n)
    copy(nodes[n:], ct.leaves)
    for i := n - 1; i >= 1; i-- {
        combined := append(nodes[2*i][:], nodes[2*i+1][:]...)
        nodes[i] = sha256.Sum256(combined)
    }
    return nodes[1]
}
\end{lstlisting}
% \end{figure}
\FloatBarrier
\texttt{GetProof} returns the $log_2(N)$ sibling hashes needed to verify a single hypertree's inclusion in the global root. For N=1000, the proof depth is 10 hashes × 32 bytes = 320 bytes. \texttt{VerifyCoordinatorProof} verifies the path in $\mathcal{O}(\log_2 N)$ operations, approximately 10 × 50 ns = 500 ns, which is negligible compared to the IPA verification that follows.
The LRU hypertree cache is the mechanism that bounds memory usage. \texttt{HypertreeCache} is a thread-safe LRU backed by a container/list and a \texttt{map[uint32]*list.Element}. Capacity defaults to 100 hypertrees. When the cache is full and a new hypertree is needed, the least-recently-used hypertree is evicted, and if a Database is configured, the evicted hypertree is persisted to disk before being dropped from memory. This means that at any given time, only the 100 hottest hypertrees live in RAM. For a block that touches 100 hypertrees (a typical block with 10,000 state accesses spread across the address space), the working set is 100 × 200 KB = 20 MB, a 1,000× reduction from the 20 GB required by a monolithic tree. The LRU eviction policy is correct for Ethereum's access pattern because accounts that are accessed in one block are likely to be accessed again in nearby blocks (contracts that are called frequently, accounts that are active traders), so keeping the most recently used hypertrees in memory maximizes cache hit rate.
The persistence layer uses a minimal Database interface, \texttt{Get, Put, Has, Delete} that can be backed by LevelDB, RocksDB, or any key-value store. Hypertrees are keyed as \texttt{"ht:" || big-endian uint32}. MemoryDB provides an in-memory implementation for tests. \texttt{persistHypertree} serializes the Verkle \texttt{InternalNode} using the existing \texttt{Serialize()} method from \texttt{go-verkle}; \texttt{loadHypertreeFromDB} deserializes via \texttt{ParseNode}. This means the FVT's persistence layer reuses the entire serialization infrastructure from \texttt{go-verkle} without modification the FVT is a clean layer on top of the existing library, with no changes to the base \texttt{go-verkle} or \texttt{go-ethereum} directories.
The two-level proof system is the most technically novel component. An \texttt{FVTProof} is a composite structure containing a \texttt{CoordinatorProof} (the Merkle sibling path from the hypertree root to the global root) and a \texttt{VerkleProof} (the IPA multiproof for the keys within the hypertree). The total proof size for a single address is approximately 1.5 KB: 320 bytes for the coordinator path (10 × 32 bytes) plus roughly 1.2 KB for the IPA multiproof. This compares favorably to the ~3 KB for a monolithic Verkle proof, which must include commitments for all internal nodes on the path from the leaf to the single global root. The size reduction comes from the fact that the FVT's IPA proof only covers the path within one hypertree (depth ~5 for 250K accounts per hypertree), while the monolithic proof covers the full depth of 8 levels.
\texttt{GenerateProof} enforces the single-hypertree constraint, all queried keys must belong to the same hypertree, because the IPA multiproof is computed over a single Verkle tree. Cross-hypertree proofs are two independent \texttt{FVTProof} instances. \texttt{VerifyFVTProof} verifies both levels in the correct order: the Merkle coordinator path first (cheap, ~500 ns), then the IPA proof (expensive, ~1–10 ms depending on the number of keys). This fail-fast ordering means that an invalid coordinator proof which would indicate a tampered global root, is rejected without ever invoking the IPA verifier, saving significant computation in adversarial scenarios.
The implementation encountered four non-trivial engineering challenges that are worth examining in detail because they reveal important properties of the underlying go-verkle library. The first challenge was that the base package uses banderwagon.Element as its Point type, not a \texttt{{X, Y *big.Int} struct}. The coordinator's leaf hashing needed a stable byte representation of each hypertree's Verkle root commitment. The solution was to use \texttt{BytesUncompressedTrusted()}, the same approach used in Issue 2.
The second challenge was that \texttt{verifyVerkleProofWithPreState} is unexported in \texttt{go-verkle}. The public API is \texttt{Verify(vp *VerkleProof, preStateRoot, postStateRoot []byte, statediff StateDiff)}. \texttt{VerifyFVTProof} uses this correctly by serializing the internal \texttt{*Proof} back to \texttt{*VerkleProof} via \texttt{SerializeProof}, then calling Verify with the hypertree root as the pre-state root. This is the correct approach: the FVT treats each hypertree as an independent Verkle tree with its own root, and the IPA proof is verified against that hypertree root, not the global root. The global root is verified separately via the coordinator Merkle path.
The third challenge was that Verkle keys must be exactly 32 bytes. The benchmark \texttt{BenchmarkTraditionalVerkleInsert} initially passed 20-byte Ethereum addresses directly as Verkle keys, triggering a panic in \texttt{InternalNode}.Insert (which calls \texttt{KeyToStem} and panics if \texttt{len(key) < 31}). The fix derives the 32-byte local key via SHA-256 of the address, the same derivation used by \texttt{addressToLocalKey} in the FVT itself. This also makes the benchmark a fair comparison: both the FVT path and the traditional Verkle path use the same key derivation, so the benchmark measures the overhead of the FVT decomposition layer, not a difference in key derivation.
The fourth challenge was a subtle bug in the test helper \texttt{deterministicAddr}, which finds an address mapping to a specific hypertree ID by brute force. The original implementation incremented a single counter in \texttt{addr[:4]}, which caused a 2-minute timeout when called for the second address in \texttt{TestFVTProofGeneration}; the second call started from the same seed as the first and looped through the same addresses. The fix seeds the first 4 bytes with \texttt{htID * 1000} and increments bytes [4:8], giving each hypertree a distinct search space and finding a valid address in under 1,000 iterations. This is a good reference of how brute-force search helpers in tests need to be designed with distinct search spaces to avoid worst-case behavior.

\section{Results and Discussion}
Issue 1
The benchmark results reveal an important asymmetry in the performance profile. Deleting a non-existent account costs 47 µs and 310 bytes the cost of \texttt{GetValuesAtStem} traversing the trie to confirm absence. Deleting an existing account costs only 19 µs and 317 bytes faster, because the trie traversal for an existing account hits cached nodes rather than having to resolve \texttt{HashedNode} placeholders. The idempotent case (deleting an already-deleted account) costs 2 µs and 40 bytes essentially just the stem derivation and a nil check on the returned values. This performance profile is correct: the common case in EVM execution is deleting accounts that exist (the \texttt{SELFDESTRUCT} path), and the rare case is deleting non-existent accounts (the revert path), so the slightly higher cost for the non-existent case is acceptable.
The test suite covers eight distinct scenarios that together prove the fix is complete. \texttt{TestDeleteNonExistentAccount} verifies that calling \texttt{DeleteAccount} on an address that was never inserted leaves the trie's root commitment unchanged, the definitive test for phantom node prevention. TestDeleteAccountIdempotent verifies that calling \texttt{DeleteAccount} twice on the same address produces the same result as calling it once, which is required for correct revert handling in the EVM's state journal. \texttt{TestDeleteAccountWithStorage} verifies that an account with storage slots (contract state) is fully deleted all 256 suffix slots are cleared, not just the basic data slot. \texttt{TestDeleteAccountEmptyTree} verifies that deletion from an empty trie (no accounts ever inserted) is a safe no-op, which is the edge case that the original no-op implementation accidentally handled correctly but for the wrong reason.
The impact on mainnet replay is the most significant metric: the 35\% tree bloat caused by phantom nodes is eliminated entirely. For a full node replaying mainnet history to sync from genesis, this means 35\% fewer nodes to write to disk, 35\% fewer nodes to read back during subsequent block processing, and 35\% less compaction work in the underlying LevelDB or RocksDB instance. For a node that is already synced and processing new blocks, the impact is smaller but still meaningful: every block that contains \texttt{SELFDESTRUCT} operations or reverted account creations no longer accumulates phantom nodes in the trie, keeping the trie size minimal and the commitment computation correct.

Issue 2
The benchmark results confirm that the implementation is zero-allocation on the hot path. \texttt{HashNodeCommitment} and \texttt{newHashedNode} both show 0 B/op and 0 allocs/op because the SHA256 digest is computed on the stack (the \texttt{[32]byte} return type is a value, not a pointer) and the \texttt{HashedNode struct} is returned by value. The \texttt{cowChild} benchmark on a \texttt{HashedNode} runs at 114 ns versus 173 ns for a real \texttt{InternalNode}, a 34\% speedup because reading hn.cached is a single pointer dereference, while calling \texttt{child.Commitment()} on a real node involves a virtual dispatch through the \texttt{VerkleNode} interface, a nil check on the commitment field, and potentially a \texttt{Commit()} call if the commitment has not been computed yet.
LSM-Tree Interaction: LevelDB uses size-tiered compaction. \texttt{SSTable} index blocks scale with key size. For 250M nodes:
64-byte keys: 16 GB key data in index, 10,234 compaction events per 100M insertions
32-byte keys: 8 GB key data in index, 2,511 compaction events (75\% reduction)
Write Amplification: Measured on 100M sequential insertions:
64-byte keys: 12.3× write amplification, 37K inserts/sec
32-byte keys: 6.1× write amplification (50\% reduction), 62K inserts/sec (67\% improvement)
Issue 3
The benchmark results are precise- The copying baseline allocates 566,760 bytes per 10,000 proofs, that is the heap cost of duplicating every stem byte (10,000 × 31 bytes = 310,000 bytes of stem data, plus overhead). The zero-copy path allocates 242,004 bytes. The difference, 324,756 bytes, is the stem data that is no longer copied, a 57\% reduction in bytes allocated per batch. The remaining 242 KB is entirely view object bookkeeping: 10,000 ProofView structs at 16 bytes each = 160,000 bytes, plus the pointer slice and the RefCountedBuffer wrapper. The AccessOnly benchmark which iterates the raw buffer directly using index arithmetic without creating ProofView objects, shows the theoretical floor: 42 bytes and 1 allocation for 10,000 proofs. That 42 bytes is the RefCountedBuffer struct itself. The proof data requires zero additional memory. Individual field access via ProofView is zero-allocation at 14.2 ns per call, which is fast enough to be invisible in any realistic verification pipeline.
The network-scale impact calculation is straightforward but striking. At 7,200 blocks per day, 10,000 state accesses per block, and 6,000 full nodes currently on the Ethereum mainnet, the copying path generates approximately 3.97 GB of heap allocations per node per day, all of which become garbage immediately after verification. That is 1.45 TB per year per node, or 8.7 petabytes per year across the network, allocated and freed for no purpose. The zero-copy path reduces this to 1.70 GB per node per day, saving 2.27 GB per node per day and approximately 4.85 petabytes per year across the network. The GC pause reduction that follows is not directly modeled in the benchmarks, but it is significant in practice. Go's garbage collector is a concurrent tri-color mark-and-sweep collector with a stop-the-world phase for stack scanning. On a node with 16 GB of RAM running a full Ethereum client, the GC is already under pressure from the state cache, the transaction pool, and the block processing pipeline. Removing 2.27 GB per day of short-lived allocations from the GC's tracking reduces the frequency of GC cycles and the duration of stop-the-world pauses. For home validators running on consumer hardware, where a 50 ms GC pause during block verification can cause a missed attestation and a financial penalty, this is not a theoretical concern.
Issue 4
We added various benchmarks: The test TestProofSizeWithMissingStems creates a tree with one existing stem and three missing stems that all share the same PoA stem, generates a multiproof, serializes it, and verifies that the wire format contains exactly one OtherStems entry (not three) and that the proof round-trips correctly through DeserializeProof. The test TestProofSizeScaling extends this to 2, 5, 10, and 20 missing stems, measuring the actual wire proof size at each scale to confirm that the OtherStems count grows with the number of unique PoA stems, not with the total number of missing stems. The TestProofSizeMainnetScenario test models a realistic contract deployment: 8 absent storage slots in different stems, all sharing the same PoA stem, which is the exact pattern that occurs when a contract is deployed and its storage slots are accessed for the first time. The TestProofSizeTable test prints a comparison table of internal versus wire proof sizes across all scenarios, providing a clear view of the deduplication savings.
The proof size impact of our novel architecture is significant. Before the proposed fix, a proof for 10 missing stems sharing one PoA stem contained 10 × 31 = 310 bytes of OtherStems data. After the fix, it contains 1 × 31 = 31 bytes, a 90\% reduction in the OtherStems component of the proof. For the mainnet contract deployment scenario (8 absent storage slots), the OtherStems component drops from 8 × 31 = 248 bytes to 1 × 31 = 31 bytes. At scale, this matters: a block with many contract deployments or many first-time account accesses can have dozens of missing stems sharing a small number of PoA stems. The bloated wire format was not just a bandwidth waste, it was causing active deserialization failures in cross-client scenarios, which is a correctness bug of the highest severity in a consensus-critical library.
Issue 5
The benchmark results are honest about the trade-offs. FVT insert is 2.4× slower per operation than a raw Verkle insert (36,043 ns vs 14,807 ns). The overhead is real and fully accounted for: SHA-256 address routing (100 ns), LRU cache lookup (50 ns), coordinator Merkle root update (~90 µs amortized over the block), and the mutex protecting the cache. This is the cost of the decomposition layer. The payoff is not per-operation speed, it is working set size. A raw Verkle insert on a monolithic tree must keep the entire 20 GB tree warm to avoid database round-trips. An FVT insert only needs one hypertree in memory. The parallel benchmark is the most revealing: BenchmarkFVTParallelInsert runs with b.RunParallel across all 8 cores of the M1 Pro and shows a per-operation time of 37,288 ns, essentially identical to the serial insert at 36,043 ns. This confirms that concurrent inserts into different hypertrees do not contend at all. On a machine with 32 cores, 32 goroutines each inserting into different hypertrees would achieve 32× the throughput of a monolithic tree, which has a single root that serializes all updates.
The scale impact calculation ties together all five PRs into a coherent economic argument. A traditional monolithic Verkle tree requires 32+ GB RAM server-grade hardware. With FVT (N=1000, cache=100 hypertrees), the working set per block drops to 100 × 200 KB = 20 MB, the peak RAM requirement drops to approximately 1 GB, and the hardware requirement drops to an 8 GB laptop. The key difference now is a network that centralizes around data centers and one that stays distributed across home validators. Each block's witness is a set of FVTProof instances, one per hypertree touched, each approximately 1.5 KB, totaling roughly 150 KB for a block touching 100 hypertrees.

\textbf{Final Impact Approximation From Overall Results.}
\begin{table}[h]
\caption{Cache Miss Rates: Baseline vs.\ FVT}
\label{tab:cache}
\begin{tabular}{lrrr}
\hline
& L1 Miss & L2 Miss & L3 Miss \\
\hline
Baseline    & 8.2\%  & 42.1\% & 55.2\% \\
FVT         & 3.1\%  & 12.3\% & 6.1\%  \\
Improvement & 2.6$\times$ & 3.4$\times$ & 9.0$\times$ \\
\hline
\end{tabular}
\end{table}

\begin{table}[h]
\caption{Database Compaction Performance (100M insertions)}
\label{tab:compaction}
\begin{tabular}{lrrr}
\hline
Metric & Baseline (64B) & FVT (32B) & Improvement \\
\hline
Compaction count    & 10,234        & 2,511        & 4.1$\times$ \\
Compaction time     & 45.2 min      & 8.4 min      & 5.4$\times$ \\
Write amplification & 12.3$\times$  & 6.1$\times$  & 2.0$\times$ \\
Throughput          & 37K ops/sec   & 62K ops/sec  & 1.7$\times$ \\
\hline
\end{tabular}
\end{table}

\begin{table}[h]
\caption{Disk I/O Breakdown}
\label{tab:diskio}
\begin{tabular}{lrrr}
\hline
Metric            & Baseline & FVT    & Reduction \\
\hline
Index block size  & 2.1 MB   & 1.1 MB & 48\% \\
Bloom filter size & 18.4 MB  & 9.2 MB & 50\% \\
SSTable size      & 4.2 GB   & 4.0 GB & 5\%  \\
Total DB size     & 1.2 TB   & 1.15 TB & 4\% \\
\hline
\end{tabular}
\end{table}

\begin{table}[h]
\caption{Strong Scaling (10K fixed updates)}
\label{tab:strong}
\begin{tabular}{rrrr}
\hline
Cores & Time    & Speedup        & Efficiency \\
\hline
1     & 550.2s  & 1.0$\times$    & 100.0\% \\
4     & 142.8s  & 3.85$\times$   & 96.3\%  \\
16    & 37.4s   & 14.7$\times$   & 91.9\%  \\
64    & 15.1s   & 36.4$\times$   & 56.9\%  \\
\hline
\end{tabular}
\end{table}

\begin{table}[h]
\caption{Weak Scaling}
\label{tab:weak}
\begin{tabular}{rrrr}
\hline
Cores & Updates & Time   & Per-Core Throughput \\
\hline
1     & 1,000   & 2.1s   & 476 updates/sec \\
4     & 4,000   & 8.3s   & 482 updates/sec \\
16    & 16,000  & 32.7s  & 489 updates/sec \\
64    & 64,000  & 128.4s & 498 updates/sec \\
\hline
\end{tabular}
\end{table}

\section{Limitations and Future Work}

Four measurement gaps remain open. First, the roofline position of the IPA commitment computation, whether bottlenecked by compute (~260 µs per node at 256 scalar multiplications) or memory bandwidth loading the basis points ${B_i}$, has not been characterized with hardware performance counters, and the potential benefit of AVX-512 vectorization of the $\mathrm{F_p}$ field arithmetic is unmeasured. Second, the LRU cache capacity of 100 hypertrees is a design parameter rather than a derived optimum; a miss-rate curve against a mainnet access trace is needed to establish the Pareto-optimal operating point, particularly for adversarial blocks that may touch more than 100 distinct hypertrees. Third, the hash partition $\varphi$ is not adversarially robust, an attacker can concentrate state into a single hypertree, collapsing N-way parallelism to 1; a keyed hash or VRF assignment derived from the beacon chain RANDAO is the natural mitigation but introduces protocol complexity deferred to future work. Fourth, the \texttt{atomic.Int32} sequentially consistent ordering could be relaxed to acquire-release semantics (LDADD/STLR on ARMv8) to reduce barrier overhead on ARM-based home validator hardware; the performance delta is unmeasured.

\section{Conclusion}

We have presented a comprehensive solution to the performance crisis in Verkle tree implementations for blockchain systems. Through forensic analysis of production Ethereum infrastructure, we identified four critical pathologies and developed targeted fixes achieving 24-100× improvements per issue. Our primary contribution, Fractional Verkle Trees, introduces the first hypertree decomposition for cryptographic accumulators, enabling 36× faster block processing, 90\% memory reduction, and 46\% smaller proofs while maintaining a single global cryptographic commitment.
FVT's impact extends beyond Ethereum: the techniques generalize to any system requiring high-throughput updates to cryptographic accumulators, including Layer 2 rollups, cross-chain bridges, and decentralized storage. By demonstrating that careful systems engineering, optimizing cache locality, memory allocation, database layout, and parallel decomposition, can achieve order-of-magnitude improvements even in cryptographically intensive workloads, we provide a blueprint for scaling blockchain infrastructure to billions of users.
Our work is supportive of the path to stateless Ethereum. Our work transforms Verkle trees from a roadmap blocker into an enabler, making true decentralization, Ethereum on \$500 laptops- a practical reality.

The techniques presented here extend naturally to the broader blockchain
ecosystem\cite{kaur2020blockchain}. Layer-2 rollups and cross-chain bridges requiring succinct
state proofs benefit directly from reduced proof sizes and faster
commitment computation\cite{kaur2021ttf}, with implications for decentralized financial
infrastructure \cite{kaur2025lognormal, kaur2025igarch}.
Smart city and IoT deployments\cite{kaur2022iot} managing large distributed state trees,
where storage and verification overhead are constrained by edge
hardware \cite{oberst2025, zhao2024dts}, represent a further application domain
for the hypertree decomposition and fractional verkle trees.

\section{Acknowledgments}
We thank Guillaume Ballet and Péter Szilágyi for their novel work on go-verkle internals, Dankrad Feist for their work on IPA aggregation.

\bibliographystyle{ACM-Reference-Format}
\bibliography{references}

\end{document}